\documentclass[11pt]{article}
\usepackage{fullpage} 
\usepackage{pstricks}
\usepackage{multicol}
\usepackage{graphicx}
\usepackage{movie15}
\usepackage{hyperref}
\usepackage[version=3]{mhchem}
\usepackage{amsmath} 
\usepackage{algpseudocode}
\usepackage{amsfonts}
\usepackage{color}
\usepackage{soul}
\usepackage{authblk}
\usepackage[margin=1.5cm,top=2.5cm,bottom=2.8cm,centering]{geometry}

\title{\begin{flushleft}\textbf{The basal level of gene expression associated with chromatin loosening shapes Waddington landscapes and controls cell differentiation \footnote{J. Mol. Biol. (2020) 432, 2253-2270. Corresponding author: denis.michel@live.fr}}\end{flushleft}}
\date{}
\author{}
\begin{document}
\maketitle
\vspace*{-2cm}
\noindent
Gilles Flouriot$ ^{1} $, Charly Jehanno$ ^{2} $, Yann Le Page$ ^{1} $, Pascale Le Goff$ ^{1} $, Benjamin Boutin$ ^{3} $ and Denis Michel$ ^{1} $\\

\noindent
$ ^{1} $ \footnotesize Univ Rennes, Inserm, EHESP, Irset UMR 1085, Rennes, France \\
$ ^{2} $ \footnotesize University of Basel, Department of Biomedicin, Basel, Switzerland \\
$ ^{3} $ \footnotesize Univ Rennes, Institut de Recherches Math\'ematiques de Rennes, France \\

\vspace*{0.1in}
\noindent
\small \textbf{Abstract}. The baseline level of transcription, which is variable and difficult to quantify, seriously complicates the normalization of comparative transcriptomic data, but its biological importance remains unappreciated. We show that this currently neglected ingredient is essential for controlling gene network multistability and therefore cellular differentiation. Basal expression is correlated to the degree of chromatin loosening measured by DNA accessibility, and systematically leads to cellular dedifferentiation as assessed by transcriptomic signatures, irrespective of the molecular and cellular tools used. Modeling gene network motifs formally involved in developmental bifurcations, reveals that the epigenetic landscapes of Waddington are restructured by the level of non specific expression, such that the attractors of progenitor and differentiated cells can be mutually exclusive. This mechanism is universal and holds beyond the particular nature of the genes involved, provided the multistable circuits are correctly described with autonomous basal expression. These results explain the relationships long established between gene expression noise, chromatin decondensation and cellular dedifferentiation, and highlight how heterochromatin maintenance is essential for preventing pathological cellular reprogramming, age-related diseases and cancer.\\

\noindent
\textbf{Keywords:} Basal gene expression, chromatin acetylation, differentiation, multistability, Waddington landscape.
\vspace*{0.2in}
%\tableofcontents
\begin{multicols}{2}

\section*{Introduction}
Data from the litterature show that basal expression, chromatin loosening and stemness, are intimately connected phenomena: \textbf{(i)} Stem cell chromatin is loosened compared to that of differentiated cells \cite{Gaspar-Maia,Moussaieff,Meshorer} and the differentiation of stem cells is accompanied by the progressive condensation of their chromatin \cite{Ugarte}. \textbf{(ii)} A high level of basal expression is notoriously important in pluripotent cells and a hallmark of stem cells, distinguishing them from their terminally differentiated counterparts \cite{Hipp,Efroni}. \textbf{(iii)} The histone mark H3K9me3 associated to closed chromatin, prevents reprogramming \cite{Matoba,Becker}. Its inhibition forbids differentiation \cite{Ugarte} whereas its forced demethylation facilitates reprogramming \cite{Antony,Wei}. \textbf{(iv)} H3K9 acetylation characterizes pluripotency and reprogramming capacity \cite{Hezroni}. \textbf{(v)} More generally, opening chromatin by inhibition of DNA methyltransferases and histone deacetylases, improves the induction of pluripotent stem cells \cite{Huangfu}. Such observations have also been reported for specialized cases of terminal differentiation. For instance, a defect of H3K9 trimethylation maintains the reprogramming capacity of CD8$ ^{+} $ lymphocytes \cite{Pace} and chromatin acetylation induces the developmental plasticity of oligodendrocyte precursors \cite{Lyssiotis}. Long before these studies, it had already been shown that cellular differentiation is associated to an overall loss of DNA accessibility, measured experimentally with DNAseI \cite{Szabo}. This impressive list of convergent observations with stem and progenitor cells can be further extended to pathological cases of dedifferentiation, notably cancer. On the one hand, cancer cell chromatin is globally decondensed, with demethylated DNA and acetylated nucleosomes, except at the level of tumor supressors. On the other hand, the aggressiveness of cancer is correlated with the degree of dedifferentiation and cell state plasticity, allowing for example cells originating from the mammary epithelium to forget their initial identity, escape hormonal control and acquire migratory properties \cite{Flouriot2014}. These systematic correlations prompted us to look for an underlying principle rooted in bifurcation circuits. The tree of cell differentiations from the egg proceeds through serial bifurcations ending with terminally differentiated cells. At each bifurcation, a pluripotent progenitor cell can give two slightly less pluripotent cell types, through genesis of bistability from a monostable system. However, the progression of bifurcations is not automatic because progenitor cells can persist indefinitely in the body, and cells can dedifferentiate and move up to pluripotent stages. We show here that the bifurcations are controlled by the level of nonspecific gene expression, itself dependent on the degree of chromatin compaction. The dedicated tool for this study is naturally the Waddington landscape, long envisioned as the ideal framework for conceptualizing cell differentiation and development. An epigenetic landscape, in the sense initiated by Waddington \cite{Waddington}, is a $ n $-dimensional potential surface shaped by the mutual compatibility or incompatibility of the concentrations of the $ n $ cellular components. Indeed, a basic principle in cellular systems is that the different macromolecules can not be present in arbitrary relative concentrations in the cell, because of the internal constraints of reticulated networks. The most direct interactions are mediated by transcription factors (TFs) and the most widely studied interaction networks are gene regulatory networks (GRN). Epigenomic and transcriptomic profiles found in large datasets emerge from such underlying circuits. Under certain conditions which are fulfilled in living systems, including positive loops and nonlinear interactions, several steady states can coexist in the landscape and the system is called multistable. In this picture of generalized interactions, the existing cellular phenotypes correspond to the possible discrete combinations, such as barcodes, of cellular components defining the bottom of the basins in the landscape. These local minima are steady states in which the different nodes of the network remain stable. All the other combinations, falling on the ridges or the sides of the mountains, are unstable and automatically pulled down by restoring forces to a basin of attraction located in the vicinity. This view illuminated our understanding of development and cellular differentiation, conceived as emerging from GRNs and biochemical circuits \cite{Kauffman,Huang1,Huang2}. In this framework, cellular differentiation is underlain by phase space translocation of gene regulatory systems from median attractors with generalized gene expression, to border attractors with selective gene expression. The former are supposed to be metastable and less resistant to fluctuations whereas the latters are classically considered stable \cite{Moris}, but experiences show that reprogramming differentiated cells generally remains possible and that conversely, multipotent cells can persist indefinitely in culture and in the organisms \cite{Moris,Nichols}. Consistent with these observations, a combination of experimental and theoretical approaches reveals here a new general principle governing cellular differentiation, in which stemness attractors dominated by median attractors, remain stable as long as the ratio of basal vs regulated transcription is high. 

\begin{center}
\includegraphics[width=8.2cm]{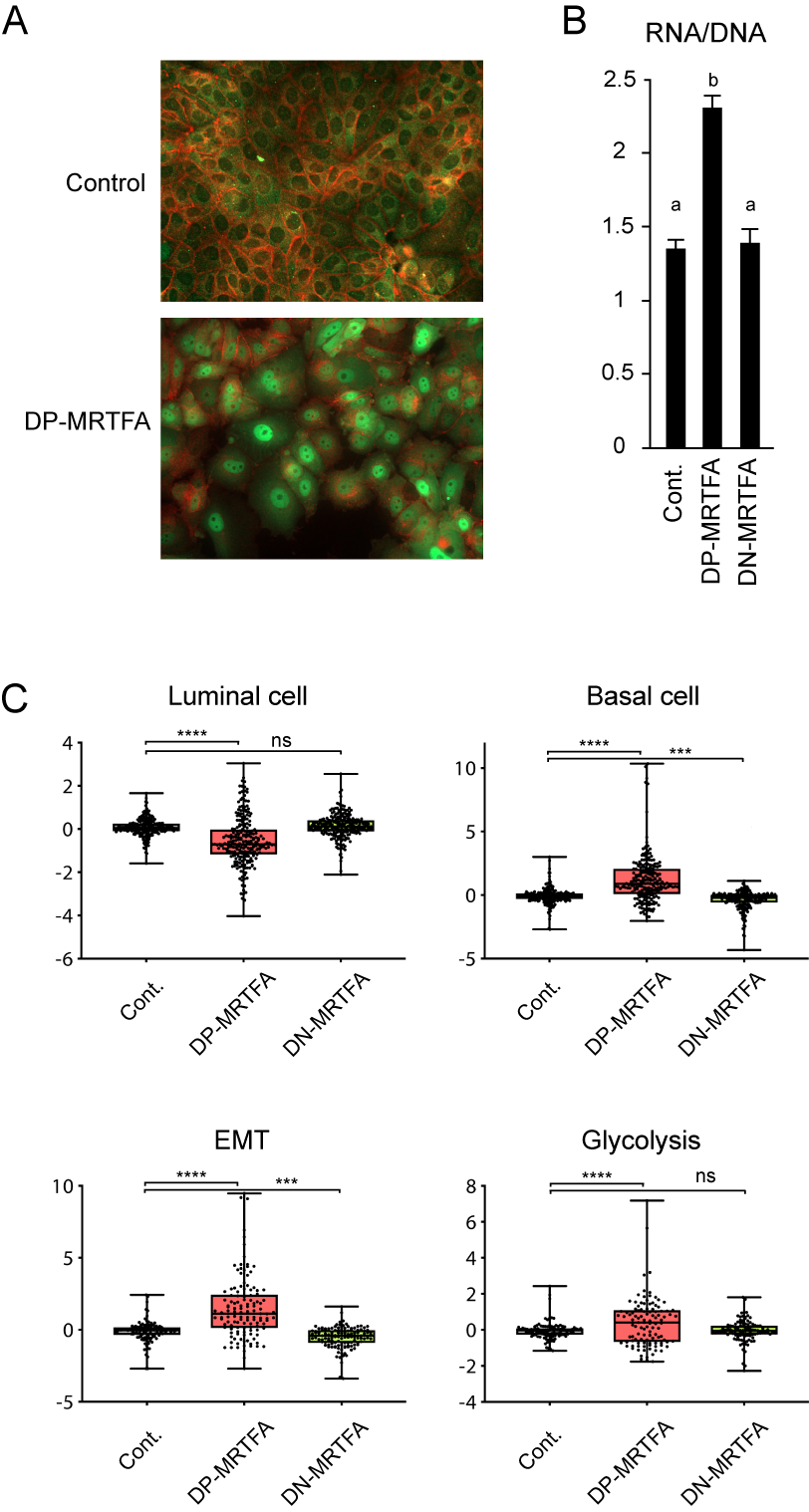} \\
\end{center}
\begin{footnotesize} \textbf{Fig. 1} Forced dedifferentiation of MCF7 cells using a constitutively active mutant of MRTFA (DP-MRTFA). (\textbf{A}) Immunofluorescence image of E-cadherin (red) and MRTFA (green) in control and DP-MRTFA-expressing MCF7 cells. DP-MRTFA causes the dismantlement of pseudoepithelial intercellular contacts, with loss of pericellular E-cadherin. (\textbf{B}) Overall transcription of control MCF7 compared with that of cells over-expressing DP-MRTFA and DN-MRTFA, determined by the ratio of cellular content in RNA over DNA. Columns with different superscripts differ significantly ($ p < 0.05 $). (\textbf{C}) Boxplot showing the log mean gene expression values obtained from RNA profiling micro-array experiment of control, DP-MRTFA and DN-MRTFA MCF7 cells of luminal, basal, EMT and glycolysis gene signatures. *$ P $-value $ < 0.05$, **$ P $-value $ <0.01 $ and ***$ P $-value $ <0.001$ with a t-test for comparisons. Error bars represent SD. \end{footnotesize}\\

\noindent
In turn, the lateral attractors with selective gene expression and characterizing terminally differentiated phenotypes, progressively deepen when lowering basal expression. 

\section*{Results}
To examine the relation between basal expression and chromatin compaction, we developped cellular/molecular instruments using mutant versions of the myocardin-related TF (MRTFA/MKL1). MRTFA has been shown to participate to a transcriptional cocktail of stemness in breast cancer cells \cite{Kim} and to erase the initial differentiation status of cells \cite{Ikeda}. But as for most biological molecules, the function of MRTFA is finely regulable, for instance by its level of expression, subcellular location and interaction partners, making it difficult to manipulate. But we showed that clear-cut functions can be imposed to MRTFA by deletion of specific interaction domains. Overexpression of a dominant positive mutant version (DP-MRTFA) devoid of cytoplasm-anchoring domain, constitutively nuclear and transcriptionally active, leads to global chromatin decondensation and induces stem cell marks such as bivalent chromatin \cite{Flouriot2014}. By contrast, an other mutant, dominant-negative (DN-MRTFA) devoid of transactivation domain, tightens chromatin and strengthens the differentiated phenotype \cite{Flouriot2014}. As shown in Fig.1A, marked phenotypic changes are induced by DP-MRTFA, with disruption of pericellular E-cadherin and of the pseudo-epithelial structure of cell monolayers.

\subsection*{Dedifferentiation, chromatin decondensation and basal expression are coupled phenomena.}
Global transcriptomic studies allowed to identify the modifications of gene expression occuring in these cells in term of signatures (Fig.1C, Table S1). DP-MRTFA caused a loss of differentiation characteristics accompanied by a clear emergence of basal cell and epithelial-to-mesenchymal transition (EMT) signatures, which characterize mammary stem cells and epithelial de-differentiation respectively. A phenomenon regularly associated with EMT: a switch of energy metabolism to glycolysis, is also obtained with DP-MRTFA. The changes induced by DN-MRTFA are globally inverse to those of DP-MRTFA, with significant decrease of the EMT and basal cell signatures, but less obvious changes of glycolysis and luminal signatures likely to be due to the fairly differentiated nature of the starting MCF7 cells. DP-MRTFA-expressing cells specifically contain large amounts of RNA per cell (mRNA + rRNA + small RNAs) (Fig.1B). This hypertranscription, which is another feature common to stem cells \cite{Turner}, could be due either to a strong increase in the specific expression of certain genes, or to a global increase of non-specific gene expression. It is technically challenging to compare gene expression between cells, because basal expression is generally unnoticed in current experimental approaches. It is ignored in transcriptome-type techniques where the results are expressed per unit mass of RNA, and/or are calibrated using gene expression supposed to be invariant between the situations. Hence, we have recourse to an old-fashioned method of nucleic acid dosage using the perchloric acid precipitation procedure \cite{Munro} to obtain accurate information on this point (Fig.1B). Because transfected DNA is packaged into nucleosomal structures similar to native chromatin \cite{Yaniv}, transient expression assays are expected to incorporate basal expression, but they have also some pitfalls: (i) It is first necessary to ensure the equivalence of transfection efficiency between the different culture conditions or the cell types to be compared. (ii) It is then necessary to find the most appropriate point of reference to quantify transcriptional changes. (i) The first requirement can in principle be satisfied by co-transfection of a neutral expression vector, provided its expression is not influenced by the more or less permissive nuclear context of each cell. A suitable internal control for this purpose is a strong promoter capable of abstracting itself from repressive contexts \cite{Koutroubas}, such as the cytomegalovirus promoter (CMV) selected here. (ii) The second point is more subtle. Transcriptional results are generally presented as "fold induction" by arbitrarily setting the uninduced condition to 1, but this presentation could introduce a bias in the interpretation of the results, because the basal (uninduced) expression level depends on the cellular context. To highlight this phenomenon, in Fig.2, the same results are presented in different manners. In these experiments, the transcriptional induction of reporter vectors is directed by either estrogen receptor (ERE) or glucocorticoid receptor response elements (GRE). ERE- and GRE-driven reporter plasmids were transfected with or without expression vectors of their respective inducers: ER$ \alpha $ with estradiol for ERE-Luc, and GR with dexamethasone for GRE-Luc. The cells tested were either human cell lines with different degrees of differentiation (Fig.2A) or MCF7 cells expressing MRTFA constructs (Fig.2B). The left histograms in Fig.2A show the results traditionally presented in fold induction, after fixing the basal expression level to 1 for each cell type. Presented in this way, the results suggest that ER$ \alpha $ and GR are less potent in dedifferentiated cells; but setting the induced level to 1, rather suggests that basal expression strongly increases in these cells. It is all the more difficult to decide which conclusion is the right one, that the compared cellular contexts largely differ. To bypass this problem of comparison, the cotransfection procedure was then applied to MCF7 cells only, which allowed to verify that the dedifferentiating construct DP-MRTFA actually increases basal expression (Fig.2B). 
\end{multicols}
\begin{center}
\includegraphics[width=17cm]{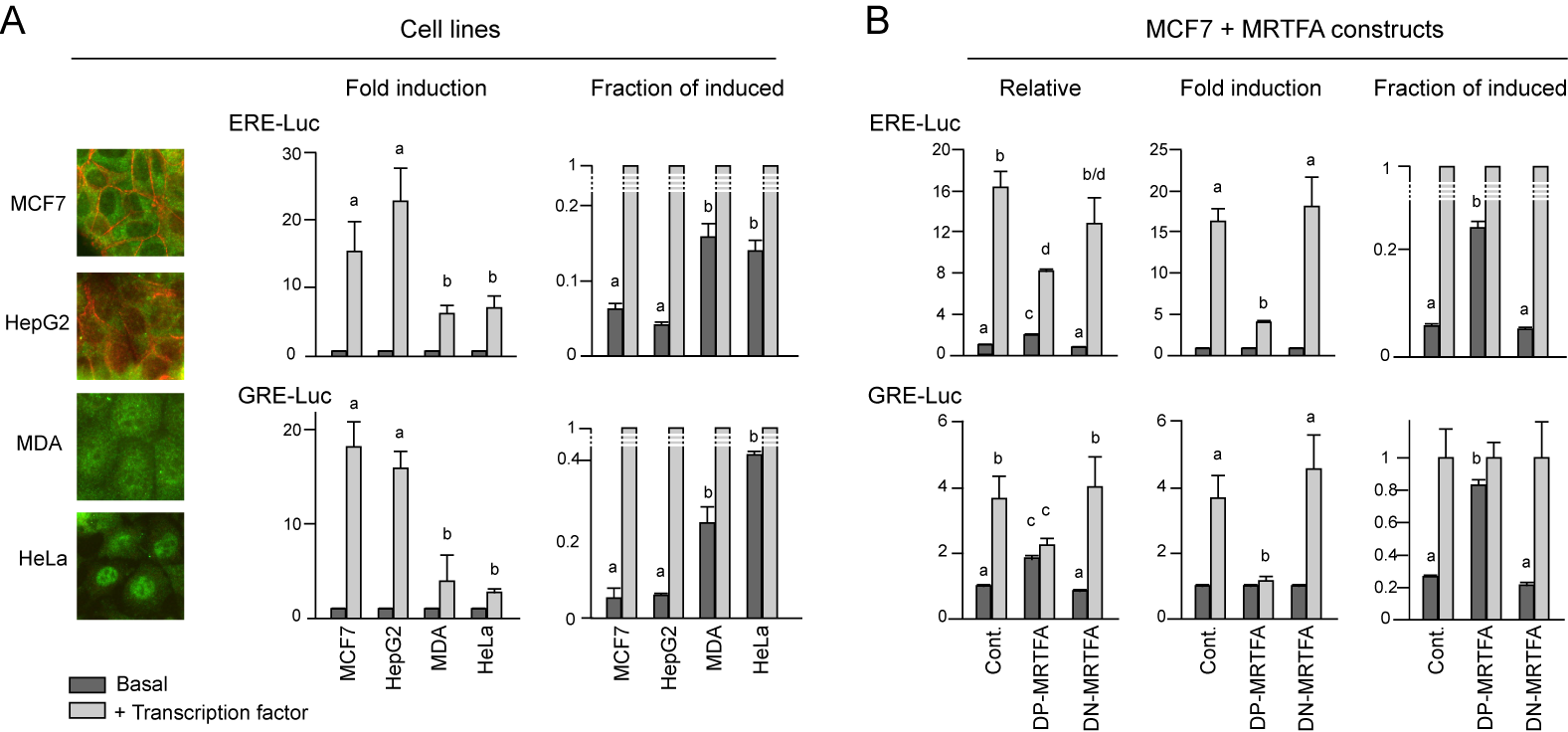} \\
\end{center}
\begin{footnotesize} \textbf{Fig. 2}. Efficient estrogen and glucocorticoid steroid receptors activation depends on the cellular state of differentiation. (\textbf{A}) Estrogen or glucocorticoid receptor-mediated transcription in cell lines characterized by their differentiated (MCF7, HepG2) or dedifferentiated (MDA, HeLa) phenotypes or (\textbf{B}) in MCF7 cells expressing or not mutant MRTFA constructs. The insets pictures in Panel A show the immunostaining of E-cadherin (red) and endogenous MRTFA (green) in the four cell lines used. Nuclear localization of MRTFA is observed in undifferentiated cell lines. All cells were transfected with the ERE-LUC or GRE-LUC reporter constructs, together with CMV-$ \beta $Gal and empty, ER, GR or MRTFA-expressing vectors. Cells were then treated with 10 nM of estradiol or 100 nM of dexamethasone. 36 hours after transfection, luciferase activities were measured and normalized with $ \beta $-galactosidase. Data correspond to the average +/- SEM of at least three separate transfection experiments. Results are expressed as either fold induction, by setting the uninduced expression level to 1, or as the fraction of the induced level, obtained by setting to 1 the induced expression. Relative expression, in the left histograms of Panel \textbf{B}, is fold changes above levels measured in control MCF7 cells. Columns with different superscripts differ significantly ($ p < 0.05 $).\end{footnotesize}
\begin{multicols}{2}

\noindent
Similar results are obtained using reporter plasmids devoid of enhancers and TATA-Box (not shown), suggesting that basal expression is weakly dependent on specific promoter sequences and could result from nonspecific interactions. The reinterpretation of results in term of variation in basal expression instead of modified induction, is unusual in the literature, which could explain why the role of basal expression is generally overlooked. Comparison of Fig.1 and Fig.2B shows that the increase vs decrease of basal expression is correlated with the tendency of dedifferentiation vs differentiation. When looking for a mechanism possibly underlying both chromatin decondensation and increase of basal expression, the most obvious candidate is the degree of chromatin acetylation. Quantification of the antagonistic chromatin marks H3K9ac and H3K9me3 shows a marked increase in acetylation in DP-MRTFA-expressing cells and inverse variations in DN-MRTFA-expressing cells (Fig.3A). As a control, the inhibitor of histone deacetylases HDAC trichostatin A (TSA), induces potent H3K9 acetylation, as expected (Fig.3B). Histone acetylation has long been shown to alleviate electrostatic interactions between nucleosomes and DNA, mechanically causing chromatin loosening, which is in turn expected to promote the accessibility of DNA to large proteins such as TFs. To test this hypothesis, we quantified the general and nonspecific accessibility of DNA with a large DNA-binding molecule, an antibody directed against double-stranded DNA. Remarkably, a significant increase of accessibility was obtained with DP-MRTFA (Fig.3). This observation is consistent with the established importance of histone acetylases in stem cells \cite{Ryall} and with the availability of their co-substrate, acetyl-CoA  \cite{Moussaieff}. Acetyl-CoA actually drops when switching the energetic metabolism from glycolysis to oxydative phosphorylation (oxphos) during differentiation \cite{Moussaieff}. Note that an inverse switch toward glycolysis is precisely observed in DP-MRTFA-expressing cells (Fig.1C).

\subsection*{Chromatin hyperacetylation is sufficient to induce some dedifferentiation characteristics}

To determine whether the triple relationship between (i) chromatin hyperacetylation, (ii) dedifferentiation and (iii) basal expression, is fortuitous or causal, we tested if chromatin acetylation caused by artificial drug treatment can induce some characteristics of DP-MRTFA-expressing cells. 

\begin{center}
\includegraphics[width=5.1cm]{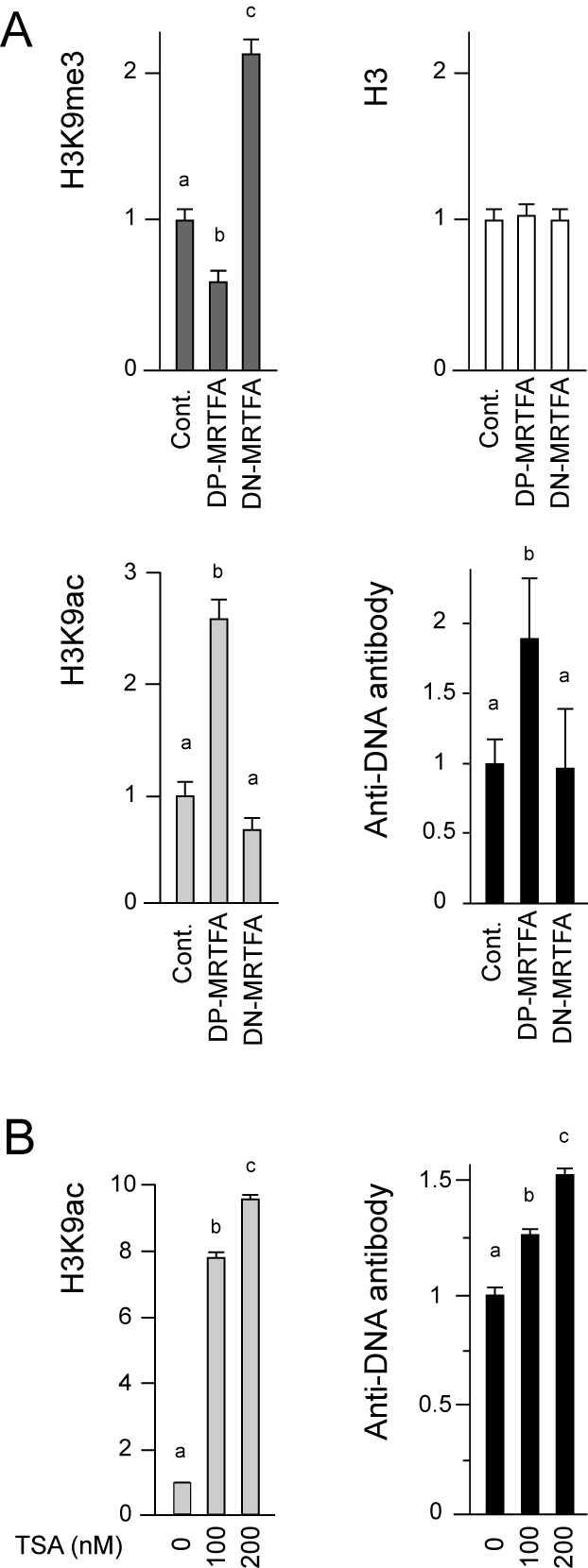} \\
\end{center}
\begin{footnotesize} \textbf{Fig. 3}. MRTFA induces global chromatin changes. (\textbf{A}) MRTFA constructs alter H3K9 trimetylation, H3K9 acetylation and DNA accessibility, evaluated by fluorescent staining using anti-dsDNA antibody, in MCF7 sub-clones 48 hours after tetracycline treatment to induce MRTFA transgenes expression. (\textbf{B}) TSA treatment induces H3K9 acetylation and DNA accessibility. Results are expressed as the fold change above levels measured in untreated MCF7 cells. Columns with different superscripts differ significantly ($ p < 0.05 $). \end{footnotesize}\\

As shown in Fig.4, mechanical  opening of chromatin using TSA, turns to be capable in itself to reproduce certain properties of DP-MRTFA cells, including a rise in basal expression (Fig.4A). In this respect, comparison of transcriptional induction by ER$ \alpha $ in the presence or absence of TSA clearly confirms the misleading character of representations in fold induction in this context.

\begin{center}
\includegraphics[width=7.8cm]{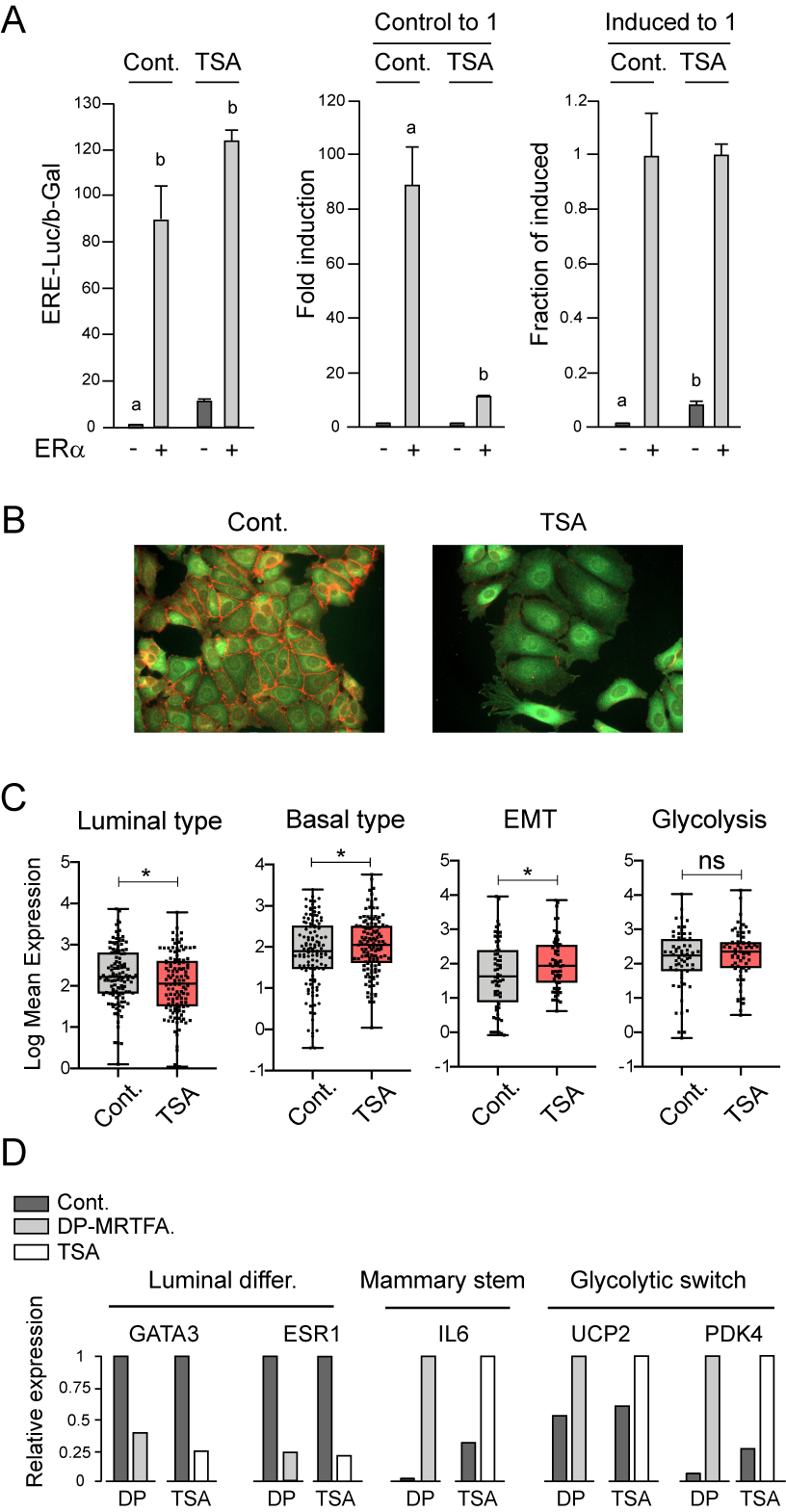} \\
\end{center}
\begin{footnotesize} \textbf{Fig. 4}. Effects of TSA on basal gene expression and transcriptomic reprogramming in MCF7 cells. (\textbf{A}) Stimulatory effect of ER$ \alpha $ on ERE-driven transcription in presence or absence of 500nM TSA. Transfections were performed as described in Fig.2. Left histograms: relative values, setting the control to 1 in absence of both TSA and ER$ \alpha $. Middle histograms: normalization of the results by setting the control to 1 (fold induction). Right histograms: normalization by setting the induced level to 1. Columns with different superscripts differ significantly ($ n $=3, $ p  < 0.05$) (\textbf{B}) Phenotypic changes induced by TSA including disruption of intercellular contacts with loss of E-cadherin (red), and a tendency of perinuclear and intranuclear accumulation of endogenous MRTFA (green). (\textbf{C}) Changes in gene expression upon TSA-treatment in MCF7 cells, analyzed from the data of \cite{Yeakley}. (\textbf{D}) Parallel effects of DP-MRTFA expression (DP) and TSA treatment on relative changes obtained in transcriptomic arrays for selected gene markers (maximal signals set to 1). \end{footnotesize}\\

When setting the control to 1, TSA seems to inhibit the activity of ER$ \alpha $ (Fig.4A, middle histograms). But when setting the induced level to 1 (right histograms), its becomes clear that this drop in fold induction may instead be due to a strong increase in basal expression. TSA treatment also induces clear phenotypic (Fig.4B) and genetic (Fig.4C) changes. 24-hour treatment with 200 nM TSA disupts cell-cell contacts and downregulates E-cadherin. TSA does not cause nuclear accumulation of endogenous MRTFA as potent as for the mutant construct DP-MRTFA, but it is however significant, particularly for larger cells, and a strong perinuclear accumulation of MRTFA is observed. A signature analysis was then conducted to determine the transcriptomic changes induced by TSA in MCF7 cells. As shown in Fig.4C, TSA has significant effects on gene expression, although less marked than with DP-MRTFA, on the decrease in luminal signature and the increase of signatures of basal cells and EMT. This transcriptomic reshaping is illustrated in Fig.4D by the changes in expression of a selection of well-identified marker genes. A decrease in GATA3 and ER$ \alpha $, the two main markers of mammary luminal differentiation, a strong increase in the mammary stem cell marker IL6 \cite{Iliopoulos,Kim}, and upregulation of genes involved in the metabolic switch to glycolysis (UCP2 and PDK4) (Fig.4D). We verified that the dedifferentiating effect of TSA can be reproduced using other HDAC inhibitors (not shown). Considering the systematic correlation observed between basal expression and cellular differentiation, it is now of interest to determine if this association is merely phenomenological or reflects a fundamental property of biological systems. To this end, we tested the influence of basal expression on particular GRNs clearly identified as regulators of cellular differentiation. 

\section*{Role of basal expression on GRN multistability}

The impact of basal expression on differentiation is tested in the framework of Waddington landscapes using two simple model systems, unidimensional and bidimensional.

\subsection*{Principle of differential GRN modeling.}
A fundamental property of living systems is the permanent renewal of all their constituents, through continuous cellular refueling with matter and energy. In this highly dynamic picture, the concentration of each constituent $ x $ results from the relative synthesis ($ S $) and removal ($ R $)

\begin{subequations}
\begin{equation} \dfrac{dx(t)}{dt}=S(t)-R(t)\end{equation}

Synthesis can itself be split into basal synthesis ($ S_{b} $), independent of the specific regulators of the considered gene, and activated synthesis ($ S_{a} $) triggered by combinations of TFs, ncRNAs and virtually all the other components of the network in an indirect manner. 

\begin{equation} S(t)= S_{b}(t)+S_{a}(t) \end{equation}

The removal of molecules also results from a basal mechanism ($ R_{b} $) generally approximated as an exponential decay, but in addition there is also the possibility of an active degradation ($ R_{a} $) by specific actors such as ubiquitin ligases for proteins.

\begin{equation} R(t)=R_{b}(t)+R_{a}(t) \end{equation}
Both basal and specific syntheses will be considered, but for removal, we will only retain, as in most studies, an exponential decay $ R(x,t)=r \ x(t) $. Basal synthesis will be reduced to a basal frequency of transcription initiation $ S_{b}(t) = b $, whereas activated synthesis $ S_{a}(t) $ is a function $ f $ combining transcription initiation frequencies $ a $ of the different TFs involved and fractional promoter occupation functions, saturable and generally nonlinear, of potentially all system's components converging to TFs. The global evolution equation of the component $ x_{j} $ thus reduces to

\begin{equation} \dfrac{dx_{j}}{dt} = b + f  \left [a_{1}, \dots , a_{n}; x_{1}, \dots , x_{n} \right ]-r \ x_{j}\end{equation}
\end{subequations} 

The function $ f $ mediating the interdependence of the different constituents of the system, impose a collective organization where only certain combinations of concentrations can remain stable. The steady states at which all constituent concentrations are mutually compatible, define the possible cell types generated by the system. Multistability is a preeminent feature of living systems and is synonymous to the capacity of differentiation. It can be obtained when (i) the system is open, subject to permanent constituent renewal, (ii) at least one positive circuit is included \cite{Kaufman} and (iii) velocities of either synthesis or removal are nonlinearly dependent on constituent concentrations. To test the effect of basal expression on the structure of the epigenetic landscape, we used abundantly documented paradigms of multistable circuits, consisting of one or two genes. Such minimalist circuits may appear ridiculously small compared with complete cellular systems, but they actually underlie real cases of bipotent progenitor differentiation. In addition, they have the practical advantage to be representable in the form of 2D and 3D landscapes.

\subsection*{Single gene circuit: the self-regulated gene encoding a dimerizable TF.}
A single autoregulated gene (Fig.5A) which is certainly one of the simplest possible circuits, is nevertheless sufficient to give rise to bistability, provided the consitions listed above are fulfilled. Despite its simplicity, this minimalist circuit is actually encountered in nature: (i) It explains for example the coexistence, in stressful conditions in a population of  Bacillus subtilis, of sporulating and vegetative cells, which is a sort of bacterial differentiation, caused by the auroregulated gene ComK \cite{Dubnau}. (ii) In vertebrates, it is involved in the vitellogenesis memory effect, evidenced in all egg-laying vertebrates tested, from fishes to birds \cite{Nicol-Benoit2}. As simple as it is, this system fulfills the three criteria listed earlier and exhibits the principle proposed here. Increasing basal expression shifts from a single attractor to two different states, of low and high gene expression, which can be regarded as a minimalist type of cellular differentiation. This one-dimensional circuit is important to consider because it can give a Waddington landscape by integration. The potential function for this unidimensional landscape can be straightly calculated by integration of the product evolution function (synthesis minus removal) \cite{Ferrell,Zhou}. Assuming a time scale separation between the DNA/TF interactions and gene expression dynamics, this minimalist gene circuit reads 

\begin{subequations}
\begin{equation} \frac{d[TF]_{\textup{tot}}}{dt}= b + a \  \frac{[TF_{2}]}{K+[TF_{2}]}-r[TF]_{\textup{tot}} \end{equation}
where $ b $ is the basal expression rate, $ a $ is the maximal rate of activated expression, $ K $ is the constant of dissociation from DNA, and $ [TF_{2}] $ is the concentration of the TF dimer, which can be defined more rigorously than using Hill functions with an exponent 2 if distinguishing the monomer, dimer and total concentrations of the TF \cite{Nicol-Benoit1}. Indeed, the gene produces the total factor in response to the dimeric factor, but the total factor concentration $ [TF]_{\textup{tot}} $ includes both monomers and dimers, such that $ [TF]_{\textup{tot}}=[TF_{1}]+2 \ [TF_{2}] $, and writing $ D $ the homodimerization constant, $ D=[TF_{2}]/[TF_{1}]^{2} $, the dimer concentration is related to its total concentration through
\begin{equation} [TF_{2}]=\left (1+4D[TF]_{\textup{tot}} - \sqrt{1+8D[TF]_{\textup{tot}}}  \right )/8D \end{equation}
When replacing $ [TF]_{\textup{tot}} $ by $ x $, the differential equation becomes
\begin{equation} \frac{dx}{dt}=b+a \ \dfrac{1+4Dx - \sqrt{1+8Dx}}{1+8KD+4Dx -\sqrt{1+8Dx}}-r x\end{equation}
\end{subequations} 

The balance between synthesis and removal terms (in short the right-hand side in the last equation Eq.(2c)), is shown in Fig.5B for the following set of parameter values: $ D=1.5 $; $ K = 0.2 $; $ a =19 $ and $r = 15$. The effect of basal expression can be straightly understood by considering the evolution function $ dx/dt $ represented in Fig.5B. Depending on the relative values of the production and removal functions, $ dx/dt $ can cross several times the null line 0, thereby yielding several possible steady states. 

\begin{center}
\includegraphics[width=8.6cm]{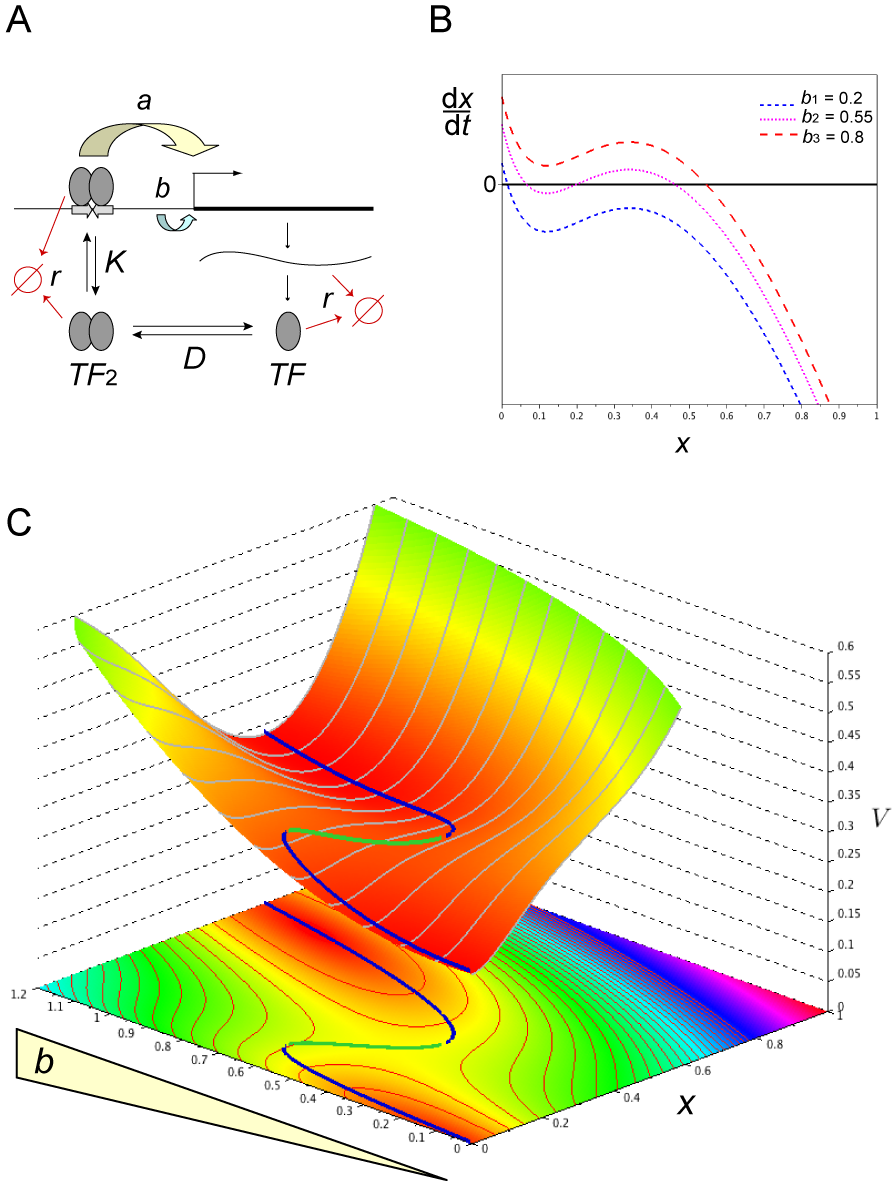} \\
\end{center}
\begin{footnotesize} \textbf{Fig. 5}. Unidimensional gene circuit. (\textbf{A}) Minimalist bistable system with a single gene activated by its own product. Gene expression is the resultant of two initiation frequencies:  $ b $, basal (independent of the TF) and $ a $, activated by the TF. (\textbf{B}) Simple mechanism to eliminate or create bistability by modifying basal expression. The number of times the function $ dx/dt $ crosses the line 0 from top to bottom, gives the number of stable steady states. A unique crossing is obtained for low and high values of $ b $ whereas two stable steady states are obtained for intermediate values of $ b $. (\textbf{C}) Evolution of the landscapes when increasing basal expression. The Waddington landscapes correspond to slices of this 3D plot (transversal curves in the $ x $ direction). \end{footnotesize}\\

To obtain such curves, degradation is generally exponential, with a flux proportional to the product concentration, while synthesis follows a saturable and sigmoidal function of the TFs. This sigmoidicity can be due to a variety of reasons, including TF dimerisation \cite{Keller,Ferrell,Nicol-Benoit1,Zhang}, or sequestration by a "poison partner" as in the next example. The resulting landscapes obtained by integration of the evolution function are shown in Fig.5C using the same parameters as mentioned earlier and for various values of $ b $.  The blue curves at the bottom of valleys stand for attractive steady state and the green one for the repulsive one. The values of $b$ for which such a repulsive steady state is present, precisely correspond to the ones for which two attractors coexist. The main parameter influencing multistability is in fact the $ a/b $ ratio, the two values $ a $ and $ b $, varying in opposite directions with the same result (Appendix B). The level of exponential degradation $ r $ is not expected to change the general principle of the mechanism presented, but to only modify the ranges of values of the $ a/b $ ratio for which the mono vs multistability domains are obtained. This is obvious in velocity equations similar to that illustrated in Fig.5B, where the production follows saturable (hyperbolic or sigmoid) functions, while degradation follows a decreasing line, cancelling out $ dx/dt $ for large values of $ x $. In its dependence in the parameter $b$, the system exhibits two connected saddle-node bifurcations. We emphasize that the parameter $b$ is not driven by any dynamic or modeling consideration. This is precisely the point to understand how its (slow) evolution may affect the whole stability properties of the (fast) system. The potential scalar function $V(b,x)$ in Fig.5C is obtained directly by integrating the opposite of the right-hand side in the evolution equation Eq.(2c):
\begin{equation}\label{eq:TFpotential}
 V = - b \ x + \dfrac{r \ x^2}{2}-a\int_{0}^{x}\!\!\tfrac{1+4D - \sqrt{1+8Dy}}{1+8KD+4Dy -\sqrt{1+8Dy}}\, dy + V_o(b) \end{equation}
This potential provides a full characterization of attractive or repulsive steady states in the following sense: the equilibria then appear respectively as (local) minimizers and maximizers of $V(b,\cdot)$. Actually, we introduce an artificial additive potential $V_o(b)$, depending only on $b$, and therefore not affecting the local minimizers and maximizers in the $x$-direction, but only their level. This allows us to obtain a readable graphical representation. The idea behind this choice is to normalize integration constants so that any of the minimizers levels are more or less independent of $b$. For simplicity, we only choose a hand-designed polynomial corrective term \[V_o(b) = 0.33 \ (b-1.2)+0.3 \ b \ (b-1.2)-0.05 \ b^2 \ (b-1.2)^2.\]
The influence of basal expression on the multistability of this toy system is well illustrated in Fig.5C, which shows a single attractor of zero expression for $ b=0.2 $, two attractors for $ b=0.55 $ and a single profound attractor for $ b=0.8 $.

\subsection*{Two gene circuit: re-modeling the celebrated GATA:PU system.}
Mutual repression has long been envisioned as a stereotyped multistability switch motif \cite{Monod}. The most popular tristable two-gene landscape is generated by a circuit with mutually repressing and self-activating genes \cite{Huang2,Moris,Zhou,Bhattacharya,Wang1,Huang2012}. Remarkably, this model corresponds to real cases of bipotent progenitor differentiation, including: (i) the balance between red and white blood cells resulting from a choice between GATA1/2 and PU.1 \cite{ZhangP}, (ii) the muscular vs vascular differentiation of somite cells determined by the Pax3:Foxc2 balance \cite{Lagha}, (iii) the ectodermal vs mesendodermal differentiation depending on the Sox2:Oct4 circuit \cite{Thomson}, and (iv) the switch from coexpression to exclusive expression of Nanog and Gata6, which directly and indirectly repress each other and specify the primitive endoderm and pluripotent epiblast respectively \cite{Plusa,Schrode}. In all these cases, upon differentiation the system evolves from a central attractor where the antagonistic genes are coexpressed, to one of two lateral attractors of exclusive gene expression. Although different genes are involved in these different tristable switches, the circuits involved can generally be reduced to stereotyped motifs \cite{Alon}, the drivers of the control of cellular behaviors and decisions. Large-scale network modeling precisely relies on the identification of these modules structuring the landscapes, by simplifying the accessory actors which do not directly modify the network topology. In turn, particular attention must be paid to the relevance of production functions for the differential analysis of the reduced networks. 
A well-established reduced motif is the GATA1/2 and PU.1 involved in blood cell differentiation. In the initial models of the GATA1/2:PU.1 circuit, self-activation and mutual repression were disconnected as depicted in Scheme 1. The variables $ x $ and $ y $ are understood as total GATA1/2 and PU.1 protein concentrations by assuming that translation is not a rate-limiting step. This modeling is widespread in the literature \cite{Zhou,Bhattacharya,Wang1} even if the basal activity is not explicitly written (set to 1) or interpreted as such.

\begin{center}
\includegraphics[width=6.5cm]{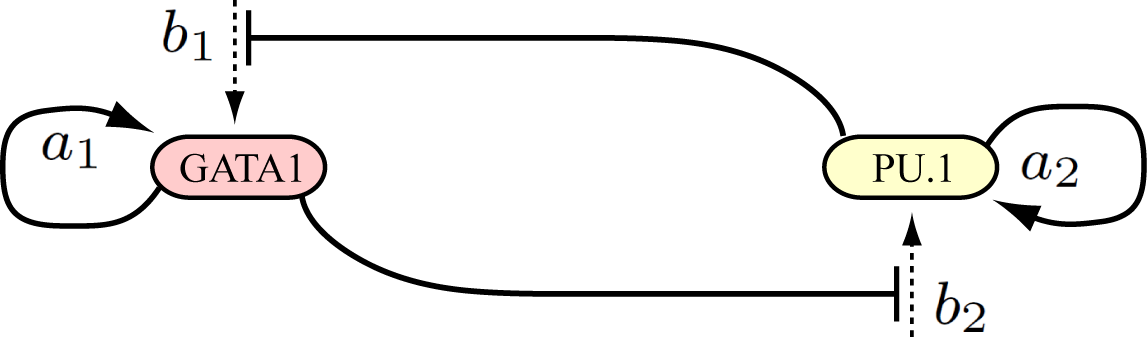} \\
\end{center}
\begin{footnotesize} \textbf{Scheme 1}. Traditional interpretation of transcriptionally independent self-stimulation and reciprocal inhibition. \\ \end{footnotesize}

The classical formulation of this scheme is
\begin{subequations} \label{E:gp}  
\begin{equation} \dfrac{dx}{dt} = a_{1} \ \dfrac{x^{n_{1}}}{K_{1}^{n_{1}}+x^{n_{1}}}+b_{1} \ \dfrac{ K_{1}^{n_{1}'}}{K_{1}^{n_{1}'}+y^{n_{1}'}}- r_{1}\ x \end{equation} \label{E:gp1}
\begin{equation} \dfrac{dy}{dt} = a_{2} \ \dfrac{y^{n_{2}}}{K_{2}^{n_{2}}+y^{n_{2}}}+b_{2} \  \dfrac{ K_{2}^{n_{2}'}}{K_{2}^{n_{2}'}+x^{n_{2}'}}- r_{2} \ y \end{equation} \label{E:gp2}
\end{subequations}

But in this scheme, basal expression would not be an independent parameter, as defined in Eq.(1d), but would be regulated by system constituents. In fact, based on the physical and functional interactions between GATA1/2 and PU.1 described in the seminal article of \cite{ZhangP}, which are recalled in Fig.6A, the mutual inhibition between the two genes does not proceed through reduction of some basal level (Scheme 1), but through preventing the self-stimulations of GATA1/2 and PU.1. This scheme thus includes genuine basal expression frequencies (Scheme 2). 

\begin{center}
\includegraphics[width=5cm]{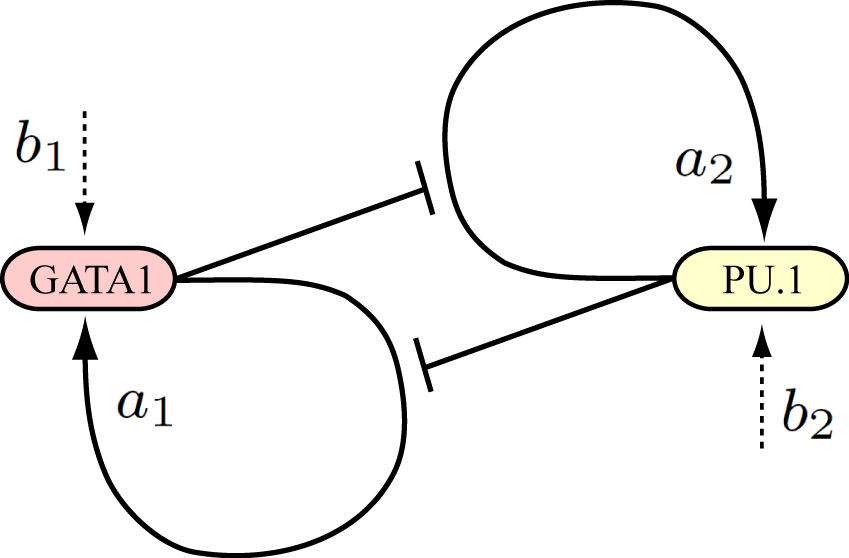} \\
\end{center}
\begin{footnotesize} \textbf{Scheme 2}. Re-interpretation of the GATA-PU motif as a reciprocal inhibition of self-stimulation, in which basal transcription frequencies are not regulated by specific factors.\\ \end{footnotesize}

A key parameter for modeling the revised mechanism of inhibited activation, is the molecular association between GATA1/2 and PU.1 (Fig.6A) \cite{ZhangP}. This association corresponds to a mutual sequestration preventing PU.1 from (i) stimulating its own gene and (ii) inhibiting the GATA1/2 gene, and vice versa (Scheme 2). The set of equations corresponding to Scheme 2 reads

\begin{subequations} \label{E:gp}  
\begin{equation} \dfrac{dx}{dt} =b_{1}+ a_{1}  \ \dfrac{x_{f}}{K_{1}+x_{f}}- r_{1} \ x \end{equation} \label{E:gp1}
\begin{equation} \dfrac{dy}{dt} =b_{2}+ a_{2} \  \dfrac{y_{f}}{K_{2}+y_{f}}- r_{2} \ y \end{equation} \label{E:gp2}
where $ x_{f} $ and $ y_{f} $ are the concentrations of molecules not mutually interacting in $ x \bullet y $ complexes. Given the time scale separation between molecular interactions (very fast) and gene expression dynamics (much slower) the free concentrations are simply given by non-differential, algebraic equations.

\begin{equation} x_{f}=x-x\bullet y \end{equation} \label{E:gp3}
\begin{equation} y_{f}=y-x\bullet y \end{equation} \label{E:gp4}

where the complex is given by
\begin{equation} x \bullet y=\frac{1}{2}\left[D+x+y-\sqrt{(D+x+y)^{2}-4xy}\right] \end{equation} \label{E:gp5}
\end{subequations}

where $ D $ is the equilibrium dimerisation constant between $ x $ and $ y $. The exponents $ n $ in the classical modeling of Eq.(4) are Hill coefficients describing molecular cooperativity, whose values are generally chosen for convenience. Arbitrarily increasing Hill's coefficients is an easy way to accentuate the relief of epigenetic landscapes, but this twist is poorly justifiable in practice in the absence of precise quantitative data. By contrast, the simple mechanism of mutual sequestration is both biologically relevant and sufficient to provide the nonlinearity necessary for multistability, whether or not the TFs work as monomers or preformed dimers. Concretely, the production functions in the differential equations are unchanged but the TF concentrations should just be replaced by their free concentrations ($ x_{f} $ and $ y_{f} $). The roles of GATA1/2 and PU.1 are supposed to be symmetrical with identical parameters for both genes ($ a_{1}=a_{2} $, $ b_{1}=b_{2} $, $ K_{1}=K_{2} $ and $ r_{1}=r_{2} $). Unlike the unidimensional evolution system Eq.(3), not any differential system in higher dimension may be described from a simple scalar-valued potential function. When a Lyapunov function exists however, it provides directly a scalar characterization of the attractive behavior of steady states and thus enables drawing of a landscape. This is the case for example for any gradient-like systems. More generally, one may try to take into account the Hamiltonian part of the dynamics, however this is not clear how to use the Hodge-Helmholtz-like decompositions to draw then a Waddington landscape \cite{Wang2,Huang2}. A general overview of landscape theories is available from \cite{ZhouLi}. An alternative approach is based on the probabilistic point of view and concerns the Freidlin-Wentzell theorem in the large deviations theory for invariant measures of stochastic convection-diffusion processes \cite{FreidlinWentzell}. In the phase space $(x,y)$, we consider many trajectories of the deterministic system Eq.(5) perturbated with a small brownian motion. These trajectories, as random variables, evolve according to a stochastic differential equation (SDE) and their probability density $p$ follows the corresponding Fokker-Planck equation, also known as Kolmogorov forward equation. This is the following partial differential equation (PDE)
\begin{equation} \label{PDE}
\dfrac{\partial p}{\partial t}+\dfrac{\partial (Xp)}{\partial x} + \dfrac{\partial (Yp)}{\partial y} = \epsilon \left(\dfrac{\partial^2 p}{\partial x^2}+\dfrac{\partial^2 p}{\partial y^2}\right)
\end{equation}
where the dynamical drift field $X(x,y),Y(x,y)$ corresponds to the respective right-hand sides in Eq.(5) and the diffusive term in the right-hand-side takes into account the random processes and thus renders the mean effect of possible noise. In large time, under some conditions on the field $(X,Y)$, most of the random trajectories of the SDE accumulates close to attractive steady states or singular trajectories of the dynamical system Eq.(5). They then evolve finally only through a fine balance between the brownian process and the deterministic dynamics. At the level of the PDE, the probability density $p$ then becomes independent of time and converges to the so-called invariant measure of the stochastic process. By simulating the convection-diffusion PDE Eq.(\ref{PDE}), we compute numerically this invariant measure to obtain its values at any point of the phase space $(x,y)$. This probability appears to be a useful tool to figure out the multistable landscape. It concentrates to high values in the neighborhood of such points but vanishes close to repulsing points. The probability in Fig.6B is obtained using a finite difference scheme to solve the PDE Eq.(\ref{PDE}) over the domain $[0,10]\times[0,10]$ in $(x,y)$ and over $[0,10]$ in the time variable. A sketch of that code is available in the appendix A. The initial data are set to a uniform density $p(t=0,x,y)=1$. The diffusion coefficient set to $\epsilon=0.025$ for the computation, is not related to a quantified information but chosen intermediately so that the equilibrium trajectories of the dynamical system Eq.(5) appear neither as singular sets (i.e. points), nor as large unidentifiable sets. For the biological parameters, we use the following set of values: $a_1=a_2=10$;  $K_1=K_2=6$, $ r_{1}=r_{2}=1 $ and $ D=1 $. In Fig.6B, the basal expression rate is set either to a low value $(b=1)$ for which bistability is present, or to a higher value $(b=4)$ with then monostability. The background color field represents the intensity of the density $p$ and the black curves figure a few deterministic trajectories, solution to the dynamical system Eq.(5) (with no brownian motion). Almost every trajectory goes in large time to one (of the two) attractive points. The computational phase space domain $[0,10]\times[0,10]$ is chosen sufficiently large so that we can prove it consists in a positively invariant domain for the dynamical system: no trajectory escape in that subset. This invariance property is useful to design convenient boundary conditions when solving the PDE Eq.(\ref{PDE}). The same strategy is used for Fig.6C. As explained earlier, the probability $p$ is then used so as to construct a landscape. More precisely the $z$ axis is defined as $ z = - q(x,b) + q_0 $, where
\begin{itemize}
\item $ q_0 $ is a constant arbitrary reference height value for the representation.
\item $ q(x,b) = max(p(x,y,b)$ for $ 0<y<10 ) $ where $ p(x,y,b) $ is the value of the solution $ p $ to the convection-diffusion PDE Eq.(\ref{PDE}) at point $ (x,y) $ for a given basal expression $ b $, computed at final time $ t=10 $ (arbitrarily chosen so as to the steady state to be reached numerically) with initial data and boundary conditions $ p=1 $. 
\end{itemize}
In some sense, this quantity $ q(x,b) $ is an easy way to represent the projection on the $ (x,b) $ plane of the probability density which currently is a scalar quantity depending on three parameters $ (x,y,b) $. We might also have represented its restriction on a two-dimensional plane set to $ y $=constant, but then the bifurcation from monostability to bistability would not have appeared so evidently. In summary, rigorous landscape treatment of the famous GATA1/2:PU.1 differentiation circuit clearly shows that a simple change in basal expression level can modify the fate of the system. For $ b=1 $, two cell types coexist (Fig.6B, left panel), whereas for $ b=4 $, a single indecise cell type exists, with equivalent coexpression of GATA1/2 and PU.1. The projection plot of Fig.6C shows the reshaping of Waddington landscape triggered by $ b $. With the set of parameters used, the transition from bistability to monostability occurs at $ b=2.54 $. 

\begin{center}
\includegraphics[width=8.8cm]{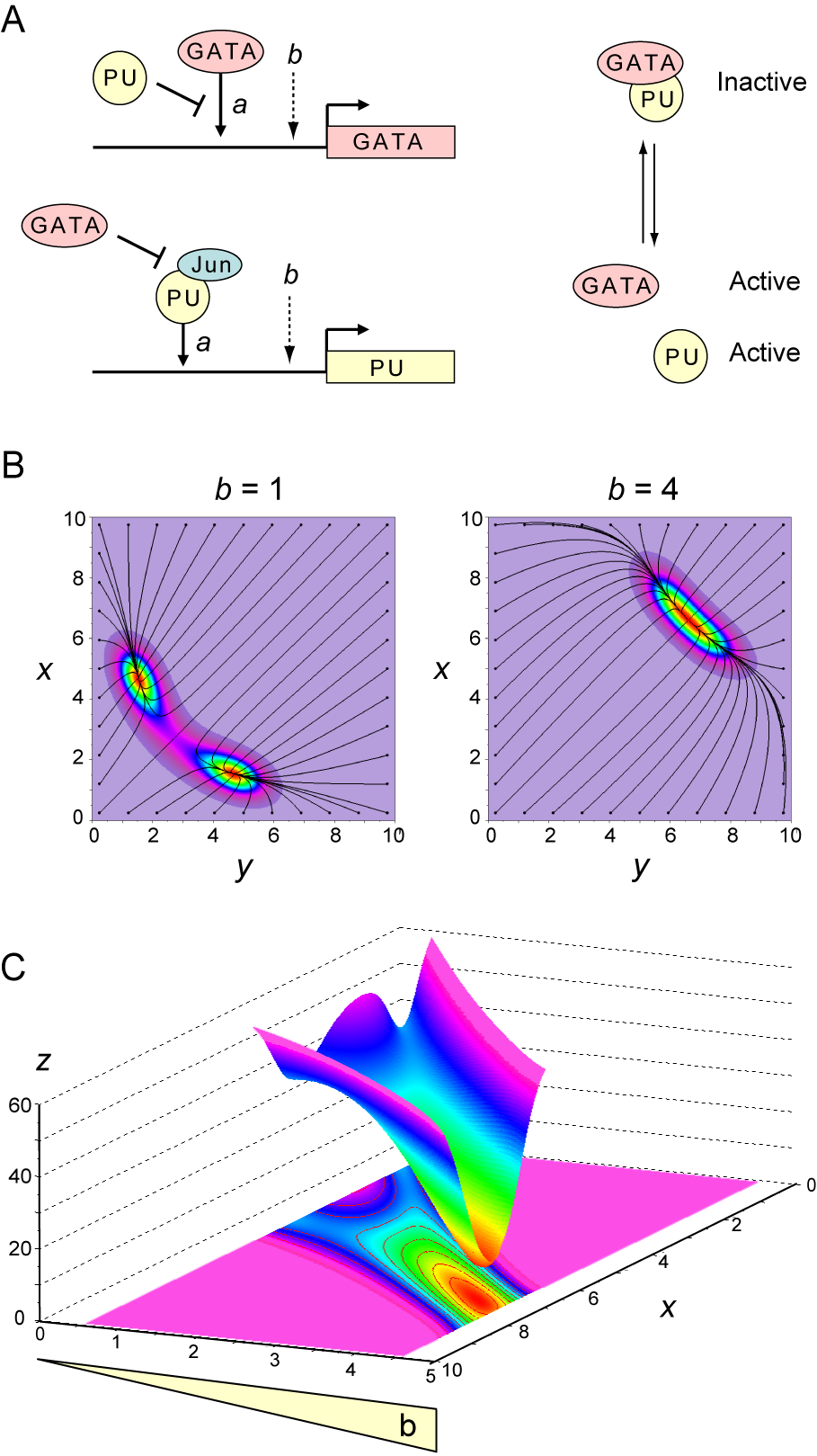} \\
\end{center}
\begin{footnotesize} \textbf{Fig. 6}. The GATA-PU system. (\textbf{A}) Summary of the molecular mechanisms involved in the self-stimulation and mutual repression of the GATA1/2 and PU.1 factors, described in the original article of \cite{ZhangP}. The physical interaction between GATA1/2 and PU.1 is the keystone of this model because only the proteins which are not mutually interacting, are active. (\textbf{B}) Waddington landscapes shaped by the GATA1/2 ($ x $): PU.1 ($ y $) circuit, according to Scheme 2, with low (left) or high (right) basal expression $ b $. (\textbf{C}) Projection of steady states on the axis of one of the variables as a function of $ b $, showing the switch from bi- to monostability when increasing $ b $, for fixed parameters $ a $ and $ K $. \end{footnotesize}\\

\subsection*{General principle of bifurcation by resolution of conflicting circuits.}
The cellular differentiation tree in development proceeds through a cascade of successive bifurcations \cite{Foster}, each one coinciding with the resolution of a gene conflict, of which a perfect example is the battle between the GATA1/2 and PU.1 genes. The present results show that a high level of basal expression alleviates the impact of their mutual repression and allows the coexpression of both antagonistic genes in progenitor cells. Lowering basal expression increases the intensity of the fight and accelerates its resolution, achieved when one of the genes loses the fight.

\section*{Model insertion in the current picture of cellular differentiation}
The present model of differentiation centered on the role of basal expression, makes it possible to weave an integrated picture combining several known properties of cellular differentiation.

\subsection*{Orientation of differentiation}
The mechanism of differentiation proposed here is attractive in that it is generic and valid for all cell lineages and applicable to various gene networks. In accordance with this model, the transdifferentiation phenomenon is expected to proceed by initial chromatin decompaction followed by recompaction. The orientation of progenitors committed to differentiate toward a particular destination attractor is supposed to result from either (i) stochastic fluctuations, favored by the low number of certain molecules such as mRNAs and allowing the cellular system to jump between adjacent attractors with a waiting time exponentially dependent on the height of the saddle point between them; or (ii) instructive exogenous inputs, like erythropoietin (EPO) and granulocyte colony-stimulating factor (G-CSF) for the red vs white blood cells, or Spemann's organizers during organogenesis, transiently altering the initial steady state. The strong heterogeneity detected in single cell transcriptomic analyses supports an important role for the first mechanism. In fact, entrusting developmental bifurcations at random is not actually a risk because (i) the proper localization of stochastically differentiated cells can be subsequently ensured by differential adhesion and cell movements \cite{Plusa}, and (ii) any imbalance in the number of cells falling in the final attractors can be corrected a posteriori by selective proliferation and/or apoptosis \cite{Plusa} to restore the appropriate partitioning of cellular masses.

\subsection*{Paradoxical decrease of regulated transcription when opening chromatin}
$ b $ and $ a $ are used here as independent parameters and only $ b $ is modified in the simulations shown here, but a mechanical link can exist between them, through which when one decreases, the other increases. In addition to the enhancers present in regulated genes, the genome contains a multitude of non-specific TF-binding sites, generally unoccupied in heterochromatin. Hence, these cryptic sites are logically exposed during dedifferentiation and can trap certain TFs, thereby reducing their free concentration and their recruitment at enhancers (thus reducing $ a $). Such a titration mechanism, which is not necessarily valid for all types of TF, has already been invoked for example to explain how the general TF TATA-binding protein (TBP) whose  concentration is limiting \cite{Manley}, puts in competition all its target genes \cite{Muldrow}. Since the chromatin of dedifferentiated cells is more accessible for proteins, as exemplified by anti-DNA antibodies (Fig.3), or DNAseI \cite{Szabo}, it seems logical that the access of TFs is also favored. The $ b/a $ ratio is therefore expected to increase upon chromatin loosening in two ways: (i) by allowing the generalized access of transcriptional machineries to a wide variety of genes (increasing $ b $) and (ii) by sequestering TFs in bulk DNA sites (decreasing $ a $). This defocusing of TFs from enhancers would therefore modify both $ b $ and $ a $, acting in concert to disimprint previous gene regulatory circuits. Simultaneous enhancer weakening and genome opening is the ideal scenario for reprogramming systems, by erasing pre-printed circuits in the face of the emergence of new actors.\\

\subsection*{Hypothetical origin of bivalent chromatin}
A particularity of histone acetylation, contrary to other histone modifications is that its role is unambiguous. For comparison, the effect of histone methylation depends on the lysine residue. Methylation of H3K4 has a permissive role and that of H3K9 methylation has a repressive effect on transcription. By contrast, histone acetylation is always permissive for TF binding regardless of the target lysine, including H3K4 in the promoter of active genes \cite{Guillemette}. Note that the methylation of H3K4 in the transcribed regions of active genes is precisely stimulated by acetylated substrates \cite{Nightingale}. A subtlety however remains to be explained: in the so-called bivalent chromatin of stem cells, acetylated H3K9 which is permissive, coexists with deacetylated and methylated enhancer H3K27 (E-H3K27) which is repressive. A hypothesis to explain this apparent paradox is based on the observation that acetylation of E-H3K27 results from the docking of histone acetyltransferases (HAT) by TFs at the level of enhancers, thereby generating a positive loop in which TF binding stimulates E-H3K27 acetylation, which in turn favors enhancer accessibility to TFs. In hyperacetylated chromatin, TF-binding motifs previously cryptic in heterochromatin become exposed and can trap TFs, reducing their free concentration and consequently their availability for binding to enhancers (supplemental model SI.2). Then, E-H3K27 methylases such as polycomb could complete the system by methylating poorly occupied E-H3K27, as verified in \cite{Ptashne}, thus precluding their reacetylation.

\subsection*{Why some genes are repressed in the context of globally open chromatin}
A long-standing enigma about cancer cell chromatin is although it is largely released, certain genes, encoding for instance tumor repressors, are closed. Similar situations are found here. In particular, TSA treatment alone is capable of both decondensing chromatin and repressing genes involved in mammary epithelial differentiation such as GATA3 and ER$ \alpha $. Candidate mechanisms to explain this, include transcriptional repressors as the zinc-fingers SNAI1 (Snail) and/or SNAI2 (Slug), known to selectively repress differentiation genes, for instance muscular \cite{Soleimani}. They are strongly upregulated in the TSA treatment by 22- and 19-fold for SNAI1 and SNAI2 respectively, and in cancer, such as during the mammary hormonal escape, where the SNAI represses ER$ \alpha $ and E-cadherin \cite{Dhasarathy}. Another excellent molecular candidate for repressing differentiation genes after chromatin opening is the polycomb system mentioned above, which could simply validate the lower occupancy of enhancers by TFs \cite{Ptashne} (supplemental model SI.2) and proceed to their closure.

\subsection*{Interplay between acetylation, metabolism and dedifferentiation}
The link between chromatin loosening and basal expression is likely to be mediated by histone acetylation, itself depending on the cellular amount of acetyl-CoA, which ultimately results from the type of energetic metabolism of the cell. This relation singularly concretizes the intimate relationship between metabolism and differentiation long anticipated by Warburg \cite{Warburg}. Warburg noticed that glycolysis is predominant in "less structured" (understand less differentiated) cells. The activity of acetylation enzymes is critically dependent on acetyl-CoA as a source of acetyl groups. Precisely, the concentration of acetyl-CoA has been shown much higher in undifferentiated cells with high glycolytic activity \cite{Moussaieff} in full agreement with the present results, including induction of glycolysis (Fig.1C) and H3K9 acetylation (Fig.3A) upon dedifferentiation.

\section*{Conclusions}
Functional correlations between chromatin loosening, dedifferentiation and basal gene expression, reflect a universal mechanism in which a decrease of basal expression systematically leads to differentiation and conversely, increasing basal expression associated to H3K9 acetylation, opens the way to reprogramming. The release of chromatin repression and the increase in non-specific gene expression naturally participate to the high entropy of the less organized undifferentiated cells. This study points out the importance of the basal expression level in GRNs, which is currently ignored in experimental as well as theoretical approaches. It is neglected in theoretical studies, as shown here for the previous modeling of the GATA1/2:PU.1 circuit, and eliminated during standardization steps in transcriptomic and epigenomic studies. The largest datasets generated by high throughput, multiplexed, single cell approaches and other modern technologies, failed to provide relevant information on the basal expression level. Reintroduction of this overlooked parameter allows to propose a unifying explanation to multiple observations including (i) the wide open chromatin of stem cells \cite{Gaspar-Maia,Moussaieff}, their generalized low level of gene expression \cite{Hipp,Efroni}, and (ii) the influence of chromatin on their differentiation \cite{Matoba,Antony,Wei,Jullien}. These different results are not only reconciled but in addition, make it possible to develop a model of cellular differentiation/dedifferentiation in the spirit of Waddington, that is to say based more on a physical principle than on particular genes. Waddington epigenetic landscapes are shown here to be structured by the level of basal gene expression. Two concurrent views of Waddington landscapes coexist: (i) a single rigid landscape specific to the genome of each organism, whose different basins correspond to the different possible cell types in the organism, or (ii) a deformable landscape whose attractors and their depth can vary with parameter adjustments. The new mechanism proposed here clearly belongs to the latter category as it predicts that the landscape is shaped by the degree of basal expression (Fig.5C and Fig.6C), in such a way that the undecided progenitor attractors remain profound as long as the cells retain their basal expression and open chromatin. This mechanism is strongly consistent with gene expression specificities of stem cells. As basal expression is progressively reduced, differentiation attractors emerge sequentially. In this gradual process of serial bifurcations, the initial commitment of totipotent cells could be triggered by a modest reduction of basal expression, while terminal differentiation of bipotent progenitors requires a strong reduction of basal expression and chromatin closure. Terminally differentiated cells with robustly imprinted circuits have a tightly packed chromatin enriched in H3K9me3, with some islets of accessibility to TFs at the level of H3K27 acetylated enhancers. The sharply partitioned chromatin of differentiated cells may ensure the persistence of well-focused specific circuits. Conversely, chromatin hyperacetylation could be responsible for a rise in basal expression and expose newly accessible binding sites for TFs, defocusing them from enhancers. A remarkable property of the present model is that differentiation and stemness attractors do not coexist at a given moment, so that stem cells cannot accidentally fall into a differentiation attractor. Conversely, the strength of established heterochromatin in differentiated cells is a powerful barrier against the risk of de-differentiation, because stem cell attractors no longer exist in that state. However, pathological or age-related loss of H3K9me3 can unlock the system and reopen the road to dedifferentiation. Hence, mechanisms ensuring the maintenance of H3K9me3 \cite{Jehanno,Becker} are essential for longevity and cancer prevention. For example, the better heterochromatinized mammary cells of formerly gestating women are less prone to cancerization, even after menopause \cite{Russo}. Although the word epigenetics was first introduced in the context of the gene networks conceived by Waddington, this term was then hijacked by researchers working on chromatin, who restricted the term epigenetics to chromatin "marks" \cite{Huang2012}. Strikingly, the present theory merges these two views by confering to chromatin epigenetics a driver role in Waddington epigenetics.

\section*{Materials and methods}
\subsection*{Plasmids, MCF7 subclones and antibodies}

The following constructs used in this study, pCR ER$ \alpha $, pSG-GR, pCR-DP-MRTFA ($ \Delta $N200), pCR DN-MRTFA ($ \Delta $C301), ERE-Luc (C3-Luc) and GRE-tk-LUC, are described in \cite{Flouriot2014}. pCMV-galactosidase, and pGL2-Basic are from Promega and pTAL-Luc from Clontech. The stably transfected MCF7 control, DP-MRTFA and DN-MRTFA subclones are described in \cite{Flouriot2014}. The following primary antibodies were used:  anti-E-cadherin (ab15148; Abcam), anti-MKL1 (sc21558; Santa Cruz Biotechnology), anti-histone H3 (E173-58; Epitomics), anti-H3K9ac (histone H3 acetylated at Lys9; ab10812; Abcam), anti-H3K9me3 H3K9me3 (histone H3 trimethylated at Lys9; ab8898, Abcam), anti-double-stranded DNA (ab27156, Abcam). Secondary antibodies conjugated to Alexa Fluor 488 or 594 were obtained from Invitrogen.

\subsection*{Cell culture, transfection, and reporter assays}
HepG2, HeLa, MCF7, MDA-MB231 (MDA), MCF7 control and constitutively expressing MRTFA constructs (T-Rex system, Invitrogen) were grown in Dulbecco's modified Eagle's medium (DMEM; Invitrogen) supplemented with 10\% fetal bovine serum (FBS, Biowest) and antibiotics (Invitrogen) at 37$ ^{\circ} $C and 5\% CO$ _{2} $ humidified atmosphere. Before all transfections and treatments, the medium was replaced with phenol red-free DMEM (Invitrogen) containing 2.5\% charcoal-stripped fetal calf serum (FCS; Biowest). Expression of the MRTFA proteins of interest was induced by a 48 h treatment of MCF7 subclones with tetracyclin. Cells were treated for 24 h when required with ligands (10 nM estradiol or dexamethasone) or ethanol (vehicle control). The treatment with trichostatin A (TSA 647925, Merck) was performed for 24 hours at 100, 200 or 500 nM. Transfection experiments were carried out exactly as previously described \cite{Flouriot2014}.

\subsection*{RNA vs DNA content}
RNA vs DNA content ratios were determined using the HClO$ _{4} $ hydrolysis method \cite{Munro}. $ 2 \times 10^{5} $ cells were cultured in 6-well plates and then dissociated after trypsinization. Macromolecules were precipitated with ice-cold 0.3N HClO$ _{4} $. After dissolution in 0.3N KOH, RNA was hydrolyzed for 1 hour at 37$ ^{\circ}$C. DNA and proteins were precipitated again in the presence of ice cold 0.3N HClO$ _{4} $ and the hydrolyzed RNA was recovered from the supernatant after centrifugation. This second pellet was dissolved in 0.6N HClO$ _{4} $, and the DNA was then hydrolyzed for 10 min at 80$ ^{\circ}$C and finally recovered after incubation for 1 h on ice and centrifugation. RNA and DNA were quantified by their absorbance at 260 nm. 

\subsection*{Immonohistochemistry}
Cells were grown on 10-mm-diameter coverslips in 24-well plates in DMEM containing 2\% charcoal-stripped FBS, treated with TSA (647925, Merck) for 24 h, and then fixed with 4\% paraformaldehyde (PFA) for 10 min and permeabilized in PBS-0.3\%Triton X-100 for 10 min. Incubation with the primary antibody (1:1000 dilution) was performed overnight at 4$ ^{\circ} $C. Secondary antibodies conjugated to Alexa Fluor were incubated for 1 h at room temperature. After washing in PBS, the cover slides were mounted in Vectashield$ \textsuperscript{\textregistered} $ medium with DAPI and images were obtained using an Imager.Z1 ApoTome AxioCam (Zeiss) epifluorescentmicroscope and processed with AxioVision Software. For each coverslip, 10 to 20 pictures were randomly taken. Pictures were visually screened in blind condition and deleted if artefactual fluorescent aggregates were present or in case of focus problems. For each picture, fluorescence values of each nucleus were obtained in an automatic manner using a homemade plugin working on Fiji. Briefly, each nucleus was identified using the DAPI labeling and after background subtraction, total fluorescence of each nucleus was extracted from the picture obtained with the fluorescent antibody. For each condition, the mean of fluorescence intensities of more than a thousand cells were calculated.

\subsection*{Transcriptomic data}
We have submitted the microarray data on MRTFA cell lines to the NCBI Gene Expression Omnibus website under accession No. GSE107924. Gene signatures were obtained from the publicly available database MsigDb Gene Set Enrichment Analysis (GSEA). The EMT and glycolysis gene signatures were obtained from the Hallmark genesets of GSEA. The luminal and basal signatures were extracted from \cite{Charafe-Jauffret} (curated gene sets). A) Comparison of the transcriptional signature of the two DP-MRTFA and DN-MRTFA clones in comparison to MCF7 control cells. B) Comparison of the four transcriptional signatures between TSA treated versus vehicle treated (Cont.) MCF7 cells. Data were obtained from by TempO-Seq targeted whole transcriptome profiling GEO accession: GSE91395 \cite{Yeakley}.

\subsection*{Acknowledgments}
We thank the Ligue R\'egionale Contre le Cancer for its sustained financial support.

\newpage
\end{multicols}
\begin{center}
\Huge{Appendices}
\end{center}
\appendix
\setcounter{equation}{0}  % reset counter 
\numberwithin{equation}{section}
\vspace*{0.2in}

\thispagestyle{empty}
\vspace{0.1cm}
\section{Algorithm for solving the convection-diffusion equation (6)}
\vspace*{0.4in}

For avoiding too many technical notations, the detailed description of the classical FFT-based 2D diffusion solver is omitted.

\vspace*{0.4in}
\begin{algorithmic}
\State $(T,x_{\max},y_{\max},\varepsilon,N) \gets (50,10,10,0.025,81)$ \Comment{Parameters}
\State $(\delta x,\delta y)\gets (x_{\max}/N,y_{\max}/N)$ \Comment{Space steps}
\State \textbf{Initialisation}
\For{$i\gets 0, N$ and $j\gets 0, N$} 
	\State $(x_i,y_j)\gets (i \delta x,j\delta y)$ \Comment{Space grid}
	\State $(X_{i,j},Y_{i,j}) \gets (X(x_i,y_j),Y(x_i,y_j))$ \Comment{Transport flow}
	\State $p_{i,j}\gets 1$\Comment{Initial probability density}
\EndFor
\State $\delta t\gets 1/(\max\!|X|/\delta x + \max\!|Y|/\delta y)$ \Comment{Time step}
\State $t \gets 0$
\While{$t<=T$}
\State $t\gets t+\delta t$
\State \textbf{convection step}
\State Extend $p$ outside $\{0,...,N\}^2$ by $0$ \Comment{Boundary conditions}
\State $\forall (i,j)\ \tilde X_{i+1/2,j} \gets \tfrac{1}{2}[X_{i+1,j}p_{i+1,j}+X_{i,j}p_{i,j}]$
\State $\forall (i,j)\ \tilde Y_{i,j+1/2} \gets \tfrac{1}{2}[Y_{i,j+1}p_{i,j+1}+Y_{i,j}p_{i,j}]$
\State $\forall (i,j)\ p_{i,j}\gets p_{i,j} - \tfrac{\delta t}{\delta x}(\tilde X_{i+1/2,j}-\tilde X_{i-1/2,j}) - \tfrac{\delta t}{\delta y}(\tilde Y_{i,j+1/2}-\tilde Y_{i,j-1/2})$
\State \textbf{diffusion step}
\State Extend $p$ outside $\{0,...,N\}^2$ by periodicity \Comment{Boundary conditions}
\State $p \gets \textsf{exp}(\varepsilon \delta t \Delta ) \ p$ \Comment{FFT-based solver}
\EndWhile
\State \textbf{return} $p$ \Comment{Final probability density}
\end{algorithmic}
\bigskip

\newpage

\section{Parsimonious model of bivalent chromatin}
\begin{multicols}{2}
An intriguing specificity of the so-called bivalent chromatin of stem cells is the presence of repressive marks (low acetylation and high methylation of enhancer H3K27, E-H3K27) and permissive marks (H3K9 acetylated and H3K4 methylated). A speculative hypothesis to explain bivalent chromatin may be based on the wide range decompaction of chromatin by acetylation of H3K9, which could be indirectly responsible for the relative closure of enhancers. Enhancers are precisely characterized, when active, by acetylated H3K27. The bivalent marks would therefore reflect a relative increase in the $ b/a $ ratio of basal to regulated transcription frequencies.
\subsection*{Model ingredients.}
In this integrated model intended to reconcile several observations with a minimum of hypotheses, stationary solutions will be obtained directly by skipping time-dependent differential equations. We will consider the existence of a single acetylation enzyme (HAT) such as CBP, capable of acetylating both H3K9 and H3K27, and a single enzyme (HDAC) capable of deacetylating them. The only difference is that in bulk chromatin, H3 lysines  (H3K9, but also possibly H3K27), can be acetylated autonomously by the HAT, whereas acetylation of E-H3K27 is assisted by TFs recruiting the HAT at their target enhancers. The second postulate is that TFs have a large number of cryptic binding sites in genomic DNA that are normally not accessible in the closed chromatin of differentiated cells, but become accessible in case of generalized decompaction. Assuming that the cellular content in $ TF $ is approximately constant, this fixation will mechanically reduce its presence on enhancers, and as a consequence decrease the maintenance of their acetylated state. For simplicity, acetylation and deacetylation of H3K9 and E-H3K27 are assumed to follow traditional Michaelis-Menten velocities with the same Michaelis constant ($ K_{\text{HAT}} $ and $ K_{\text{HDAC}} $). The enzymes are supposed to bind to both lysines, but to be significantly sequestrated by H3K9 only, considering that the E-H3K27 sites are restricted to enhancers, so that for a diffusing enzyme, the accessible concentration of H3K9 is much higher than that of E-H3K27.
\subsection*{Fraction of acetylated H3K9.}
H3K9 acetylation and methylation are mutually exclusive marks, but we will consider here that the dynamics of H3K9 acetylation/deacetylation is fast enough compared to that of methylation, to allow considering only the unmethylated fraction of H3K9.
Let us define dimensionless Michaelis constants weighted by the substrate concentrations, with
$$  \text{[H3K9]tot}=N $$
$$  K_{A}= K_{\text{HAT}}/N $$ and 
$$  K_{D}= K_{\text{HDAC}}/N $$ 

The maximal velocity of acetylation is 
$$ V_{A}=c_{A}\text{[HAT]tot [acetyl-CoA]} $$
where $ c_{A} $ is the catalytic rate, and the velocity of deacetylation is $$ V_{D}=c_{D}\text{[HDAC]tot} $$. 

Writing $ V $ the sum of maximal velocities $ V= V_{A}+V_{D} $, we define fractional maximal velocities  $$ \theta=\dfrac{V_{A}}{V} $$ and $$ 1-\theta=\dfrac{V_{D}}{V} $$
Using this nomenclature, the fraction of acetylated H3K9 (written $ \rho $) and that of deacetylated H3K9 ($ 1- \rho $) are given by the traditional zero-order mechanism formulated in Table 1.

\begin{table*}
\centering
\caption{Acetylated and deacetylated fractions of unmethylated H3K9.}
\bigskip
\label{tab:1}
\begin{tabular}{l|c}

\hline
& \\
$ \dfrac{\text{[H3K9]ac}}{\text{[H3K9]tot}} $ & $  \rho= \dfrac{(1-\theta)(K_{A}+1)+\theta (K_{D}-1)-\sqrt{((1-\theta)(K_{A}+1)+\theta (K_{D}-1))^{2}-4\theta(1-2\theta)K_{D}} }{2(1-2 \theta)} $  \\
&  \\
$ \dfrac{\text{[H3K9]}}{\text{[H3K9]tot}} $  & $ 1-\rho $  \\
\hline
\end{tabular}
\end{table*}

\subsection*{TF sequestration by H3K9-acetylated chromatin.}
Simple statistics and systematic sequencing have shown that consensual and near-consensual DNA-binding sites for most TFs are widespread in the genome, but that only a few of them correspond to genuine regulatory elements or enhancers. ChIP-seq experiments confirmed that these cryptic putative binding sites are generally not occupied in native chromatin, suggesting that their accessibility is prevented by chromatin closure. Hence, we will postulate here that chromatin loosening by acetylation could render the cryptic sites accessible. In random sequences, cryptic sites are distributed on average every $ n $ nucleosomes. Their concentration is $ R= $ [H3K9ac]/$ n = \rho N/n$, $ n $ being about 20 for a consensus sequence of 6 base pairs in a random sequence. If the TF binds to these sites with an average dissociation constant $ K =k_{d}/k_{a} $, a fraction of the TF (of constant total concentration $ F $) will be sequestrated, yielding only a residual free concentration $ f $, such that

\begin{subequations}
\begin{equation} R =\dfrac{(F-f)(K+f)}{f} \end{equation}
giving
\begin{equation} f =\frac{1}{2}\left(F-K-R+\sqrt{(F+K+R)^{2} -4RF} \right )\end{equation}
\end{subequations}

\subsection*{E-H3K27 acetylation.}
The knowledge of the free concentrations of TF and enzymes finally allows to predict the acetylation status of E-H3K27. Contrary to that of H3K9, E-H3K27 acetylation is supposed to necessitate previous TF binding. In turn TF binding is poorly efficient in absence of E-H3K27ac and is then greatly facilitated by E-H3K27ac. The rates used for the different reactions are listed in Table 2. 

\begin{table*}
\centering
\caption{E-H3K27 acetylation/deacetylation cycle assuming that the HAT is recruited by the TF. $ f $ is the concentration of free TF (not sequestrated by cryptic elements) $ k_{a}f $ and $ k_{d} $ are the pseudo-first order and first order rate constants of TF binding to and dissociation from acetylated chromatin respectively. The binding of TF to the deacetylated enhancers is supposed possible, but with a constant $ h $ lower than $ k_{a} $.}

\bigskip
\label{tab:2}
\begin{tabular}{l|c}
\hline
& \\
Transition  &  Rate  \\
&  \\
\hline
&  \\
E-H3K27 $ \rightarrow $ E-H3K27-F   & $ hf $  \\
&  \\
E-H3K27-F $ \rightarrow $ E-H3K27  & $ k_{d} $  \\
&   \\
E-H3K27-F $ \rightarrow $ E-H3K27ac-F & $ \dfrac{c_{A}\textup{[free HAT]}}{K_{\text{HAT}}} = \dfrac{V}{N}\dfrac{ \theta}{K_{A}+1-\rho} $  \\
 \\
E-H3K27ac-F $ \rightarrow $ E-H3K27ac  & $ k_{d} $  \\
 \\
E-H3K27ac $ \rightarrow $ E-H3K27ac-F  & $ k_{a}f $  \\
 \\
E-H3K27ac $ \rightarrow $ E-H3K27 & $\dfrac{c_{D}\textup{[free HDAC]}}{K_{\text{HDAC}}} =\dfrac{V}{N}\dfrac{ 1-\theta}{K_{D}+ \rho} $  \\
&  \\
\hline
\end{tabular}
\end{table*}

In absence of specific data, we will arbitrarily assume that the catalytic rate and Michaelis constant of the HAT are the same for H3K9 in absence of $ F $ and for E-H3K27 in presence of TF. In turn, enzymatic sequestration is assumed to be caused by H3K9 only considering the minor contribution of the enhancers in the genomes. Using the first-order or pseudo-first order rates of Table 2, the stationary probabilities of E-H3K27 acetylation is 

\begin{equation} \begin{split}& P(H3K927ac)=\\& \dfrac{1}{1+ \dfrac{ k_{d}\left[N (k_{d}+hf)(K_{A}+1-\rho)+V \theta \right](1-\theta)}{hf\left[N (k_{d}+k_{a}f)(K_{D}+\rho)+V(1-\theta) \right]\theta } } \end{split} \end{equation}

Replacing in this equation $ f $ by its value given in Eq.(B.1) and $ \rho $ by its value given in Table 1, allows to express $ P(H3K927ac) $ as a function of the variable $ \theta $ only. As represented in Fig.B.1, E-H3K27 can be largely deacetylated in spite of an overall hyperacetylation in the cell. In other words, a global increase of acetylases activity can simultaneously open bulk chromatin and alter enhancers, leading to an increase of $ b $ and a decrease of $ a $ for $ \theta >0.5 $ (Fig.B.1). This model is minimalist in that it is based only on the competition between bulk histone acetylation and E-H3K27 acetylation, and recourses to very few ingredients in a field, chromatin epigenetics, which involves a lot of molecular actors. Other correlations and amplification phenomena, not incoporated here are naturally expected to complete the picture. For instance, methylation maintains non-acetylated lysines in non-acetylable form by competition, thereby locking the system. Conversely the histone variant H2A.Z, whose profile is parallel to that of H3K27ac at the level of enhancers, is also an important player in enhancer functions by causing nucleosomal depletion \cite{Pandey}. Certain marks are clearly correlated: H3K9 acetylation is associated to H3K4 methylation, and is closely related to DNA methylation in multiple ways including:

\begin{center}
\includegraphics[width=7cm]{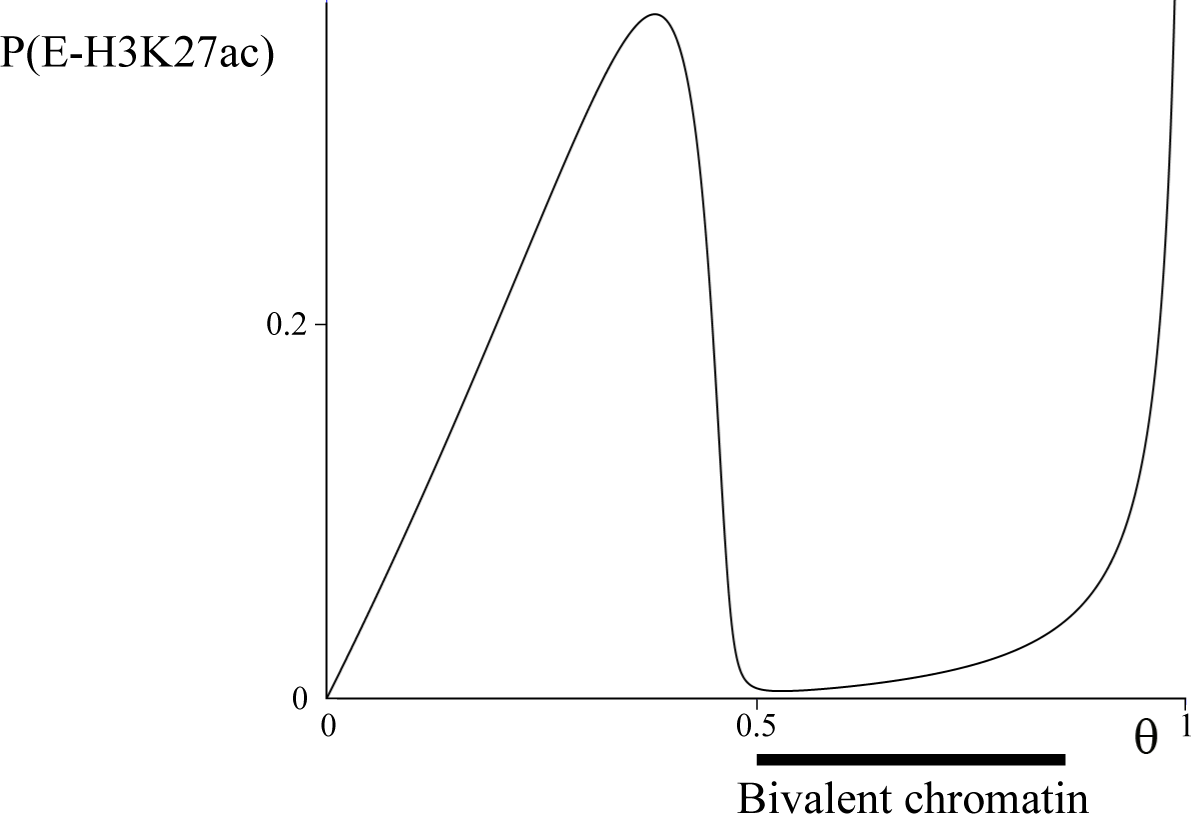} \\
\end{center}
\begin{footnotesize} \textbf{Fig. B.1}. Curve drawn to Eq.(B.2) for the set of parameters $ k_{a}=k_{d}=1 $, $ h=0.001 $, $ K_{a}=K_{d}=0.1 $, $ N=3000 $, $ n=10 $ and $ F=100 $  \end{footnotesize}\\

 (i) the presence of methyl-cytosine binding protein MBD1 in H3K9 methyl transferase complexes like SETDB1 and CAF1, (ii) the recruitment of HDACs by methylated DNA-bound MeCP2 and conversely (iii) the recruitment of a DNMT by H3K9me3-bound HP1. H3K9 methylation can stabilize chromatin in a non-acetylable form in post-mitotic differentiated cells. It should be noted in this respect that H3K9me3 is particularly persistent and constitutes the main lock against the risk of de-differentiation and reprogramming \cite{Matoba}. From the Waddington-type view, the developmental selection of the genes to close in the course of differentiation, proceeds more by absence of expression than by active repression. Repressive machineries like polycomb complexes, which suppress the expression of many genes in embryonic stem cells \cite{Turner}, could only ratify preexisting low transcription states, as suggested in \cite{Ptashne}, thereby locking selectively the genes which have already lost their Waddington-type fight in antagonistic genetic circuits. The relative contributions of basal vs regulated expression ($ b/a $) are the main regulator of the balance of cellular dedifferentiation/differentiation in the present model. A speculative and a general scenario depicted in Fig.B.2, can thus be proposed around this central core, which connects several results of cellular biology, from metabolism to multistability, which is the fundamental hallmark of differentiation.\\
 
\begin{center}
\includegraphics[width=2.7cm]{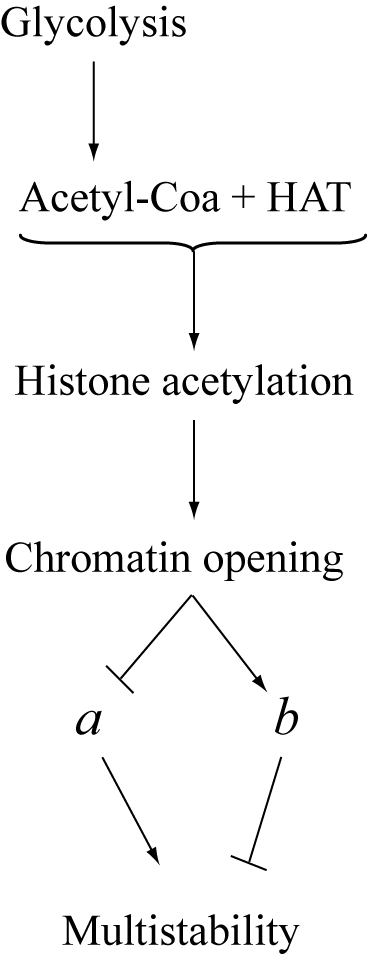} \\
\end{center}
\begin{footnotesize} \textbf{Fig. B.2}. Hypothetical scheme connecting metabolism to inhibition of multistability, equivalent to dedifferentiation.\\ \end{footnotesize}
\end{multicols}

\end{document}